\documentclass[12pt]{article}
\usepackage[cp1251,cp866]{inputenc}
\usepackage[T2A]{fontenc}
\usepackage{upref}
\usepackage{xspace}
\usepackage{enumerate}
\usepackage{amsmath,amsfonts,amssymb,cite}
\usepackage{amsthm}
\usepackage{amsxtra}
\usepackage{amscd}
\usepackage{array}

\newcommand{\textprime}{$\mathsurround0pt'$}
\newcommand{\bysame}{\hspace{1sp}\raise2pt\hbox to3em{\hrulefill}}

\def\head#1\endhead{\section*{\ignorespaces#1\unskip}}
\def\subhead#1\endsubhead{\subsection*{\ignorespaces#1\unskip}}

\numberwithin{equation}{section}

\def\bey#1\eey{\begin{equation}#1\end{equation}}
\def\ben#1\een{\begin{equation*}#1\end{equation*}}
\def\bst#1\est{\begin{split}#1\end{split}}

\newcommand\1[1]{\expandafter\bey\bst#1\est\eey}
\newcommand\2[1]{\expandafter\ben\bst#1\est\een}

\newcommand{\rom}[1]{\textrm{\mdseries\upshape#1}}

\let\cal\mathcal
\let\Bbb\mathbb

\def\qq{\qquad}

\def\qu{\quad}

\def\Det{\operatorname{Det}}
\def\eff{\operatorname{eff}}

\def\const{\operatorname{const}}
\def\cond{\operatorname{cond}}
\def\free{\operatorname{free}}
\def\sign{\operatorname{sign}}

\date{}

\begin{document}

\renewcommand{\appendixname}{Appendix}
\renewcommand{\refname}{References}
\renewcommand{\proofname}{Proof}


\title
{{\bf The Functional Integration and}\\
 {\bf the Two-Point Correlation Function of}\\
 {\bf  the One-Dimensional Bose Gas}\\
 {\bf in the Harmonic Potential}
}

\author{
$${}$$
$${}$$
\\
{\bf N.~M.~Bogoliubov
          \!\footnote{\emph{E-mail}: bogoliub@pdmi.ras.ru} ,
          C.~Malyshev
          \!\footnote{\emph{E-mail}: malyshev@pdmi.ras.ru}
}
\\
$${}$$
\\
{\it V. A. Steklov Institute of Mathematics,}\\
{\it St.-Petersburg Department,}\\
{\it Fontanka 27, St.-Petersburg, 191023, RUSSIA}\\
$${}$$
}

\maketitle

\begin{abstract}
\noindent
A quantum field-theoretical model which describes spatially non-homogeneous
one-dimensional non-relativistic repulsive Bose gas in an external
harmonic potential is considered. We calculate the two-point thermal
correlation function of the Bose gas in the framework of the functional
integration approach. The calculations are done in the coordinate
representation. A method  of successive integration over the ``high-energy''
functional variables first and then over the ``low-energy'' ones is used.
The effective action functional for the low-energy variables is
calculated in one loop approximation. The functional integral representation
for the correlation function is obtained in terms of the
low-energy variables, and is estimated by means of the stationary
phase approximation. The asymptotics of the correlation function is studied
in the limit when the temperature is going to zero while the volume occupied
by non-homogeneous Bose gas infinitely increases. It is demonstrated that
the behaviour of the thermal correlation function in the limit described
is power-like, and it is governed by the critical exponent which depends on
the spatial and thermal arguments.
\end{abstract}

\vskip0.5cm

\leftline{2000 \emph{Mathematics Subject Classification}.
            Primary 81S40; secondary 42C10}

\vskip0.2cm

\leftline{\emph{Key words and phrases}.
            Bose gas; functional integration; correlation function}

\newpage

\head
\S1.~Introduction
\endhead

A recent burst of interest to theory of the Bose gas is caused by
experimental realization of Bose condensation in the ultra-cold
vapors of alcali metals confined in the magneto-optical traps
\cite{1, 2}.
In particular, it became possible to study the Bose condensation in the
systems, which are effectively two-dimensional or quasi one-dimensional
\cite{3, 4}. Here a partial localization along one or two
directions in three-dimensional system is achieved by making the
level spacing of the trapping potential in the corresponding
directions larger than the energies of individual atoms. The field
models which describe the Bose particles with the delta-like
inter-particle coupling confined by an external harmonic potential
provide a good approximation for a theoretical approach to the
experimental situation \cite{5}. The theory of non-ideal Bose gas
attracts traditionally not only physicists but also mathematicians
\rom{[6--8]}. For a translationally invariant homogeneous Bose
gas, the field model in question corresponds to a quantum
nonlinear Schr\"odinger equation, which admits an exact solution
in the one-dimensional case
\cite{9, 10}. This fact allows to obtain closed expressions for
the correlation functions
\cite{11, 12}.

In real physical systems the interest to a transition from
three-dimensional to one-dimensional behaviour is caused by the
fact that an effective density of atoms in the one-dimensional
Bose gas can be either high or low depending on the parameters of
the system \cite{13}. Here a low density implies a strong coupling
between the particles \cite{14, 15}, while a high density
corresponds to a weak interaction.

The present paper continues the series of the papers
\rom{[16--20]} devoted to investigation of the correlation
functions of the Bose gas with weak repulsive inter-particle
coupling in the presence of an external harmonic potential. Since
there is no exact solutions in the case of an external potential,
in the present paper, as well as in Refs. \rom{[16--20]}, the
method of functional integration \rom{[21--24]} is accepted for
investigation of the correlation functions. In the present paper
we develop the results of the paper \cite{20}, where, in a
distinction with \rom{[16--19]}, an explicit dependence of the
correlation functions on the imaginary time for non-zero
temperature was taken into account. It will be demonstrated that
the presence of the external potential (of the {\it trap}) results
in a change of the asymptotical behaviour of the two-point
correlation function in comparison to a translationally invariant
case. It will be demonstrated that the observed change happens in
the range of temperatures, which are comparable with inverse of
the characteristic length of the trap provided this length tends
to infinity.

The paper is organized as follows. Section \S1 has an introductory
character. A description of the one-dimensional model of
non-relativistic Bose field in question, as well as a summary of the
method of the functional integration, are given in Section \S2. An
approach to approximate investigation of the functional integrals
is also presented in Section \S2. This approach is based on a
successive integration first over the high-energy over-condensate
excitations and then over the variables, which correspond to the
low-energy quasi-particles. Besides, in this Section we give a
derivation of one loop effective action for the low excited
quasi-condensate fields, and the corresponding energy spectrum of
the low lying excitations is obtained. The method of stationary
phase is used in \S3 for an estimation of the functional integral,
which expresses the two-point thermal correlation function of the
non-homogeneous Bose gas. The method of asymptotical estimation of
the correlators, which is used in the present paper, was proposed in
\cite{25}, where the asymptotical behaviour of two-point Green
functions of the homogeneous Bose gas was investigated for the spatial
dimensionalities one, two, and three. In the
present paper it is demonstrated that the stationary phase method
\cite{25} admits a generalization for the spatially
non-homogeneous Bose gas in the external potential also. The
asymptotics of two-point correlation functions of non-homogeneous
Bose gas are obtained in Section \S4 both at nonzero and zero
temperatures. A short discussion of the results of the present
paper is given in Section \S5.

\head
\S2.~The effective action and the Thomas--Fermi approximation
\endhead

\subhead{1.~The partition function}
\endsubhead
In the present paper we shall consider one-dimensional Bose gas
described by the Hamiltonian $\widehat H$ defined on the real axis
$\Bbb R\ni x$:
$$
\aligned
\widehat H&=\int \left\{ \widehat \psi^{\dagger}\left(x\right) {\cal H}
\widehat \psi \left( x\right) + \frac g2 \widehat \psi^{\dagger}\left(x\right)
\widehat\psi^{\dagger }\left( x\right) \widehat \psi \left( x\right)
\widehat \psi\left(x\right) \right\} dx,
\\
{\cal H}&\equiv \frac{-\hbar^2}{2m}\frac{\partial^2}{\partial x^2}
                            -\mu+V\left( x\right),
\endaligned
\eqno{(1)}
$$
where $\widehat \psi^{\dagger }(x)$ and $\widehat \psi (x)$ are
operator-valued fields, which describe creation and annihilation
of the quasi-particles over the Fock vacuum $|0\rangle$. The fields
$\widehat \psi^{\dagger }(x)$ and $\widehat \psi (x)$ are subjected
to the commutation relation
$$
\widehat \psi (x)\widehat\psi^{\dagger }(x^{\prime })-
\widehat \psi^{\dagger }(x^{\prime })\widehat \psi (x)=\delta (x-x^{\prime })
$$
($\widehat \psi^{\dagger }(x)$ and $\widehat \psi (x)$ are
mutually commutative), and ${\cal H}$ is the ``single-particle''
Hamiltonian. The following notations are used in equation (1): $m$
is the mass of the Bose particles, $\mu $ is the chemical
potential, $g$ is the coupling constant which corresponds to the
weak repulsion (i.~e., $g>0$), and the external confining
potential is chosen in the form of the harmonic potential
$V(x)\equiv \frac m2\Omega ^2x^2$.

Let us begin with the investigation of the partition function
$Z$. It can be represented in the form of the functional integral
\rom{[21--24]}:
$$
Z=\int e^{S[\psi ,\bar \psi ]}{\cal D}\psi {\cal D}\bar\psi,
\eqno{(2)}
$$
where $S[\psi,\bar \psi ]$ is the action functional of the system in question:
\begin{align*}
S[\psi ,\bar \psi ]
=\int\limits_0^\beta d\tau \int dx
&\Big(\bar \psi (x,\tau )
\Big(\frac \partial {\partial \tau }-{\cal H}\Big)\psi (x,\tau )
\\
&\qu-\frac g2\bar \psi (x,\tau ) \bar \psi (x,\tau )\psi (x,\tau)
\psi (x,\tau )\Big).
\tag{3}
\end{align*}
The domain of the functional integration in (2) is given by the space
of complex-valued functions $\bar \psi (x,\tau )$, $\psi (x,\tau )$
depending on two arguments: $x\in\Bbb R$ and $\tau\in [0, \beta]$.
With regard to the first argument $x$, the functions $\bar\psi (x,\tau )$,
$\psi (x,\tau )$ belong to the space of quadratically integrable
functions $L_2(\Bbb R)$, while they are finite and periodic with the period
$\beta =(k_BT)^{-1}$ with regard to the imaginary time $\tau$ ($k_B$
is the Boltzmann constant, and $T$ is an absolute temperature).
The variables $\bar\psi$, $\psi $ are the independent variables
of the functional integration \cite{21}, and ${\cal D}\psi
{\cal D}\bar \psi $ is the functional integration measure.

At sufficiently low temperatures we can expect to assume that each
of the variables $\bar\psi (x,\tau )$, $\psi (x,\tau )$ is given
by two constituents. One of them, $\bar\psi_o (x,\tau )$, $\psi_o
(x,\tau )$, corresponds to {\it quasi-condensate}, while another
one --- to the high-energy thermal (i.~e., over-condensate)
excitations $\bar\psi_e (x,\tau )$, $\psi_e (x,\tau )$:
$$
\psi (x,\tau )=\psi_o(x,\tau )+\psi _e(x,\tau ),\quad
\bar\psi (x,\tau )=\bar\psi_o(x,\tau )+\bar\psi _e(x,\tau).
\eqno{(4)}
$$
It should be stressed that here and below just the quasi-condensate
is assumed, since a true Bose condensate does not exist in
one-dimensional system \cite{6}. In the exactly solvable case, the
existence of the quasi-condensate implies that a non-trivial
vacuum state (i.e., a ground state) exists. The quasi-condensate
variables $\bar\psi_o (x,\tau )$, $\psi_o (x,\tau )$ can also be
represented as the sums of the constituents:
$$
\psi _o(x,\tau )=\psi _o(x)+\xi (x,\tau ),\quad
\bar\psi _o(x,\tau )=\bar\psi _o(x)+\bar\xi (x,\tau ),
\eqno{(5)}
$$
where the field $\psi_o (x)$ describes the ground state of the model
at zero temperature, while the field $\xi (x,\tau )$ describes the low
lying excited particles. Let us require the variables in (4) to be
orthogonal in the following sense:
$$
 \int \psi_o(x,\tau )\bar\psi_e(x,\tau ) dx=
 \int \bar  \psi_o(x,\tau )\psi_e(x,\tau) dx=0.
$$
As a result, the integration measure ${\cal D}\psi {\cal D}\bar \psi$
will be replaced by ${\cal D}\psi_o{\cal D}\bar \psi_o{\cal D}\psi_e
{\cal D}\bar \psi_e$.

To investigate the functional integral (2), we shall perform a
successive integration over the fields $\bar\psi, \psi$. First, we
shall integrate over the high-energy constituents, and then --
over the low-energy ones (see.~(4)) \cite{21, 24}. At a second
step, it is preferable to pass to new functional variables, which
describe an observable ``low-energy'' physics \cite{21, 24} in a
more adequate way. After the substitution of the expansion (4)
into the action (3) we shall take into account in $S$ only the
terms up to quadratic in $\bar\psi_e$, $\psi_e$. This means that
we are making an approximation in which the over-condensate
quasi-particles do not couple with each other. In this case, it is
possible to integrate over the thermal fluctuations $\bar\psi_e
(x,\tau )$, $\psi_e (x,\tau )$ in a closed form and thus to arrive
to an effective action functional $S_{\eff}[\psi_o, \bar \psi_o]$
depending only on the quasi-condensate variables $\psi_o, \bar
\psi_o$:
$$
S_{\eff}[\psi _o,\bar \psi _o]=\ln \int
e^{\widetilde S[\psi_o+\psi_e ,\bar\psi_o+\bar\psi_e ]}
{\cal D}\psi_e{\cal D}\bar \psi_e,
\eqno{(6)}
$$
where the tilde in $\widetilde S$ implies that ``self-coupling''
of the fields $\bar\psi_e (x,\tau )$, $\psi_e (x,\tau )$ is
excluded. With respect to equation (6), the partition function $Z$
of the model takes an approximate form:
$$
Z \approx \int e^{S_{\eff}[\psi_o,\bar \psi_o]}{\cal D}\psi_o{\cal D}
     \bar \psi_o.
\eqno{(7)}
$$

Let us consider the derivation of the effective action $S_{\eff}[\psi_o,
\bar \psi_o]$ (6) in more details. The splitting (4) allows to derive
$S_{\eff}[\psi_o, \bar \psi_o]$ in the framework of the field-theoretical
approach of the loop expansion \cite{26, 27}. We substitute (4) into the
initial action $S[\psi ,\bar\psi ]$ (3) and then pass from $S$ to the action
$\widetilde S$, which is given by three terms:
$$
\widetilde S = S_{\cond}+S_{\free}+S_{\rom{int}}.
\eqno{(8)}
$$
In (8), $S_{\cond}$ is the action functional of the condensate
quasi-particles, which corresponds to a tree approximation \cite{26, 27}:
\begin{align*}
S_{\cond}[\psi_o,\bar \psi_o] \equiv \int\limits_0^\beta d\tau \int dx
&\Big\{ \bar\psi_o(x,\tau )\widehat K_{+}\psi_o(x,\tau )
\\
&\qu-\frac g2\bar\psi_o(x,\tau )\bar\psi_o(x,\tau )
\psi_o(x,\tau )\psi_o(x,\tau)\Big\}.
\tag{9}
\end{align*}
At the chosen approximation, the action for the over-condensate
excitations $S_{\free}$ takes the form:
$$
S_{\free}[\psi_e,\bar \psi_e]\equiv \frac 12\int\limits_0^\beta d\tau \int dx
\bigl(\bar \psi_e,\psi_e\bigr) \widehat G^{-1}
\begin{pmatrix} \psi_e \\
                \bar \psi_e
\end{pmatrix}.
\eqno{(10)}
$$
Eventually, $S_{\rom{int}}$ is given by the part of the total action
functional, which describes the coupling of the quasi-condensate to the
over-condensate excitations:
\begin{align*}
S_{\rom{int}}[\psi_o,\bar\psi_o,\psi_e,\bar\psi_e] \equiv \int\limits_0^\beta
d\tau \int dx
&\Big\{\bar\psi_e(x,\tau )
[\widehat{K}_{+}-g \bar\psi_o\psi_o]\psi_o(x,\tau )
\\
&\qu+ \psi _e(x,\tau)[\widehat{K}_{-}-g\bar\psi_o
\psi_o] \bar\psi_o(x,\tau )\Big\}.
\tag{11}
\end{align*}
In formulas (9)--(11), we have defined the differential operators
$\widehat{K}_{\pm }\equiv\pm \partial/ {\partial \tau }-{\cal H}$
(here the Hamiltonian ${\cal H}$ is defined in (1))
and the matrix-differential operator $\widehat{ G}^{-1}$:
$$
\widehat{ G}^{-1} \equiv \widehat{G}_0^{-1}-\widehat{\Sigma },
\eqno{(12)}
$$
where
$$
\widehat{G}_0^{-1} \equiv
\begin{pmatrix}   
                  \widehat{K}_{+} & 0 \\
                  0 & \widehat{K}_{-}
\end{pmatrix},   
\quad
\widehat{\Sigma }\equiv \widehat{\Sigma }[\psi_o,\bar \psi_o] = g
\begin{pmatrix}    
                2\bar\psi_o \psi _o & \psi_o^2 \\
                (\bar \psi_o)^2 & 2\bar\psi_o\psi_o
\end{pmatrix}.   
$$

In this approach it is appropriate to apply the stationary phase
method to the functional integral (6). To this end, let us choose
$\bar \psi_o,\psi_o$ as the stationarity points of the functional
$S_{\cond}$ (9), which are defined by the extremum condition
$\delta (S_{\cond}[\psi_o,\bar \psi_o])=0$. The corresponding equations
are taking the form of the Gross--Pitaevskii-type equations \cite{28}:
$$
\aligned
\Big(\frac \partial {\partial \tau }+\frac{\hbar ^2}{2m}\frac{\partial^2}
{\partial x^2}+\mu -V(x)\Big) \psi_o-g (\bar\psi_o \psi_o) \psi_o &=0,   
\\     
\Big(-\frac \partial {\partial \tau }+\frac{\hbar ^2}{2m}\frac{\partial^2}
{\partial x^2}+\mu -V(x)\Big)
\bar \psi_o-g (\bar\psi_o\psi_o)\bar \psi_o &=0.  
\endaligned
\eqno{(13)}
$$
The contribution of the action $S_{\rom{int}}$ (11) drops out from (8)
since $\bar\psi_o, \psi_o$ are chosen to be solutions of equations (13).
Therefore the dynamics of $\psi_e,\bar \psi_e$ is described, in the
leading approximation, by the action $S_{\free}$ (10). The latter
depends on $\bar \psi_o,\psi_o$ non-trivially through the matrix
of the self-energy parts $\widehat{\Sigma }$, which enters
into $\widehat{G}^{-1}$ (12). The {\it Thomas--Fermi approximation}
is essentially used in the present paper in order to determine
the stationarity points $\bar\psi_o, \psi_o$. This approximation
consists in neglection of the kinetic term
$\frac{\hbar ^2}{2m}\frac{\partial^2} {\partial x^2}$ in equations
(13) \cite{5, 28}. The Thomas--Fermi approximation is valid
for the systems containing a sufficiently large number of particles,
and it is widely used in the theoretical approaches to description
of the Bose condensation in the magneto-optical traps \cite{5, 28}.
The following condensate solution can be obtained provided only
$\tau$-independent solutions of (13) are allowed:
$$
\bar\psi_o\psi _o=\rho _{TF}(x;\mu )\equiv \frac 1g\left(\mu
-V(x)\right) \Theta\left( \mu -V(x)\right),
\eqno{(14)}
$$
where $\Theta$ is the Heavyside function. Now the integration in (6)
with respect to $\psi_e$, $\bar\psi_e$ is Gaussian.
This leads \cite{21} to the one loop effective action, which
takes the following form in terms of the variables $\psi_o$,
$\bar\psi_o$:
$$
S_{\eff}[\psi _o,\bar \psi_o]\equiv S_{\cond}[\psi_o,\bar \psi_o]-
\frac 12 \ln \Det (\widehat{G}^{-1}).
\eqno{(15)}
$$
Here $\widehat{G}^{-1}$ is the matrix operator (12), and
$\psi_o$, $\bar\psi_o$ have a sense of the new variables, and
their dynamics is governed by the action (15).

In order to assign a meaning to the final expression for the effective
action (15), it is necessary to regularize the determinant
$\Det (\widehat{G}^{-1})$. In our case, the operator $\widehat{G}^{-1}$
is already written as a $2\times 2$-matrix Dyson equation (12),
where the entries of $\widehat{\Sigma }[\psi _o,\bar \psi _o]$
play the role of the normal ($\Sigma_{1 1}$ $=$ $\Sigma_{2 2}$) and
anomalous ($\Sigma_{1 2}$, $\Sigma_{2 1}$) self-energy parts.
The Dyson equation (12) defines the matrix $\widehat{G}$, where the
entries have the meaning of the Green functions of the fields
$\bar\psi_e, \psi_e$. The matrix $\widehat{G}$ arises as a formal inverse
of the operator $\widehat{G}^{-1}$:
$$
\widehat{G}=\left( \widehat{G}_0^{-1}-\widehat{\Sigma }\right) ^{-1}.
\eqno{(16)}
$$
The matrix operator $\widehat{G}^{-1}$ (12) can formally be
diagonalized by means of the famous N.N.Bo\-go\-liu\-bov's
$(u,v)$-transform \cite{6}. The corresponding equations, which
describe the unknown coefficient-functions $u,\,v$, result in a
compatibility requirement, which determines, in turn, the
quasi-classical spectrum of the elementary excitations \cite{28}.

With regard to our purposes, it is appropriate to represent
$\widehat{G}^{-1}$ as follows:
$$
\widehat{G}^{-1}=\widehat{G}_0^{-1}-\widehat{\Sigma }\equiv
\widehat{{\cal G}}^{-1}-(\widehat{\Sigma }-2g\rho _{TF}(x;\mu )\widehat I),
\eqno{(17)}
$$
where $\widehat I$ is the unit matrix of the size $2\times 2$, and
the matrix $\widehat{{\cal G}}^{-1}$ is defined as
$$
\widehat{{\cal G}}^{-1}\equiv
\begin{pmatrix}
         \widehat{K}_{+}-2g\rho _{TF}(x;\mu ) & 0 \\
         0 & \widehat{K}_{-}-2g\rho _{TF}(x;\mu )
\end{pmatrix}  
\equiv
\begin{pmatrix}
              {\cal K}_{+} & 0 \\
              0 & {\cal K_{-}}
\end{pmatrix}.   
\eqno{(18)}
$$
Here $\rho_{TF}(x; \mu )$ is the solution (14), and equation
(17) implies that we simply added and subtracted
$2g\rho_{TF}(x;\mu )$ on the principle diagonal of the matrix
operator $\widehat{G}^{-1}$. A formal inverse of the operator
$\widehat{{\cal G}}^{-1}$ can be found from the following equation,
which defines the Green functions ${\cal G}_{\pm }$:
$$
\begin{pmatrix}
           {\cal K}_{+} & 0 \\
           0 & {\cal K_{-}}
\end{pmatrix}
\begin{pmatrix}
          {\cal G}_{+} & 0 \\
          0 & {\cal G}_{-}
\end{pmatrix} 
=\delta (x-x^{\prime })\delta (\tau -\tau ^{\prime })\widehat{I}.
$$

Using the relation $\ln\,\Det={\rom {Tr}}\,\ln$, one gets:
\begin{equation*}
-\frac 12\ln\,\Det \widehat{G}^{-1}
=-\frac 12\text{\rm Tr\, ln}\,
\left( \widehat I-\widehat{{\cal G}}\bigl(\widehat{\Sigma }
-2g\rho _{TF}(x;\mu )\widehat I\bigr)\right) -\frac 12\text{\rm ln}\,\Det
\begin{pmatrix}
            {\cal K}_{+} & 0 \\
            0 & {\cal K_{-}}
\end{pmatrix}.
\tag{19}
\end{equation*}
The first term in Right Hand Side of (19) is free from divergencies.
Let us consider the determinant of the matrix-differential operator
in Right Hand Side of (19). The operators ${\cal K}_{\pm }$
can be written in the form:
$$
{\cal K}_{\pm }\equiv \pm \frac \partial {\partial \tau }+
\frac{\hbar ^2}{2m}\frac{\partial ^2}{\partial x^2}+|V(x)-\mu |.
\eqno{(20)}
$$
Let us denote the eigenvalues of the operators ${\cal K}_{\pm }$ as
$\pm i\omega_B -\lambda _n$, where $\omega_B$ are the bosonic
Matsubara frequencies, and $\lambda _n$ are the energy levels
(which are labeled by the multi-index $n$) of the operator
$-\frac{\hbar ^2}{2m}\frac{\partial ^2} {\partial x^2}-|V(x)-\mu |$.
The regular part of the logarithm of the determinant in equation (19)
has the sense of the free energy $\widetilde{F}_{nc}$ of the ideal gas
of the over-condensate excitations:
$$
\widetilde{F}_{nc}(\mu )\equiv \frac 1{2\beta }\text{\rm ln}\, \Det
\begin{pmatrix}
           {\cal K}_{+} & 0 \\
           0 & {\cal K_{-}}
\end{pmatrix}  
=\frac 1\beta \sum\limits_n\text{\rm ln}\,
\Big( 2\sinh \frac{\beta \lambda_n}2\Big),
$$
where the regularized values of the determinants of the operators
${\cal K}_{\pm}$ can be obtained, for instance, by means of
zeta-regularization approach \cite{27}. Then, in the leading order in
$g$, one gets:
\begin{align*}
&-\frac 12 \text{\rm ln}\, \Det \widehat{G}^{-1}
\\
&\qq
\approx-\beta \widetilde{F}_{nc}(\mu)
+g\int\limits_0^\beta d\tau \int dx\bigr( {\cal G}_{+}(x,\tau;x,\tau )
+{\cal G}_{-}(x,\tau;x,\tau )\bigl)
\bigl(\bar\psi_o\psi_o-\rho_{TF}(x;\mu )\bigr)
\\&\qq\equiv -\beta F_{nc}(\mu )+2g\int\limits_0^\beta d\tau \int dx\rho_{nc}(x)
\bar\psi_o\psi_o.
\tag*{\normalsize(21)}
\end{align*}
\unskip Here $F_{nc}$ is the free energy of the non-ideal gas of
the over-condensate quasi-particles, and the last term in (21)
describes a coupling of the over-condensate quasi-particles with
the condensate. The density of the over-condensate quasi-particles
is $\rho_{nc}(x)\equiv -{\cal G}_{\pm }(x,\tau ;x,\tau )$, and it
depends only on the spatial coordinate $x$. At very low
temperatures and sufficiently far from the boundary of the domain
occupied by the condensate, the quantity $\rho_{nc}(x)$ can
approximately be replaced by $\rho_{nc}(0)$, since ${\cal G}_{\pm
}(x, \tau; x, \tau )$ is almost constant over a considerable part
of the condensate \cite{29}. Eventually, one obtains
\begin{align*}
S_{\eff}[\psi_o,\bar \psi _o]
=-\beta F_{nc}(\mu )
+\!\int\limits_0^\beta \!d\tau \!\int \!dx
\Big\{ \bar \psi _o(x,\tau )\Big(\frac{\hbar ^2}{2m}
\frac{\partial ^2}{\partial x^2}+\Lambda -V(x)\Big)\psi_o(x,\tau)&
\\
-\frac g2\bar \psi _o(x,\tau )\bar \psi_o(x,\tau)\psi_o(x,\tau )
\psi_o(x,\tau )&\Big\},
\tag{22}
\end{align*}
where $\Lambda =\mu -2g\rho _{nc}(0)$ is the renormalized chemical
potential. We shall consider $S_{\eff}$ (22) as one loop effective
action, where the thermal corrections over the ``classical''
background are taken into account. The ``classical'' background
corresponds to the solution (14). It should be noticed that the
described derivation of the effective action does not depend on
spatial dimensionality, and therefore it is valid for two and
three dimensions also.

In the effective action (22) it is appropriate to pass to the new variables,
namely the density $\rho (x,\tau )$ and the phase $\varphi (x,\tau )$
of the field $\psi_o(x,\tau )$ \cite{21}:
$$
\psi _o(x,\tau )=\sqrt{\rho (x,\tau )}e^{i\varphi (x,\tau )},
\quad
\bar \psi_o(x,\tau )=\sqrt{\rho (x,\tau )}e^{-i\varphi (x,\tau )}.
\eqno{(23)}
$$
We shall consider $\rho (x,\tau )$ and $\varphi (x,\tau )$
as two new independent real-valued variables of the functional integration.
Now the integration measure ${\cal D}\bar \psi_o {\cal D}\psi _o$
is replaced by the measure ${\cal D}\rho {\cal D}\varphi$.
In these new variables the effective action (22) takes the form:
\begin{align*}
S_{\eff}[\rho ,\varphi ]
&=-\beta F_{nc}(\mu )+i\int_0^\beta d\tau \int
dx\Big\{ \rho \partial _\tau \varphi +\frac{\hbar ^2}{2m}\partial _x
(\rho \partial _x\varphi)\Big\}
\\
&\qu+\int_0^\beta d\tau \int dx\Big\{\frac{\hbar ^2}{2m}(\sqrt{\rho}\,
\partial_x^2\sqrt{\rho }-\rho (\partial_x\varphi )^2) +
(\Lambda -V) \rho -\frac g2\rho ^2\Big\}.
\tag{24}
\end{align*}
Here and below we denote the partial derivatives of the first order
over $\tau $ and~$x$ as $\partial _\tau $ and $\partial _x$,
respectively, whereas the partial derivatives of the second order
--- as $\partial _\tau ^2$ and $\partial _x^2$. Notice
that equations (22) and (24) remain correct at $V=0$, so that the
renormalized chemical potential is still given by the equation
$\Lambda =\mu -2g\rho _{nc}(0)$, where $\rho _{nc}(0)$ is the
``bare'' \cite{21} density of the condensate.

\subhead{2.~The excitations spectrum}
\endsubhead
Let us consider the problem of determination of the spectrum of the
low-energy quasi-particles. We shall apply the stationary phase
approximation to the integral (7), where the effective action
is given by (22). The corresponding stationarity point is determined
from the extremum condition $\delta \bigl(S_{\eff}[\rho,
\varphi ]\bigr)=0$, which is equivalent to the couple of the
Gross--Pitaevskii equations:
$$
\aligned
 i\partial _\tau \varphi +\frac{\hbar ^2}{2m}\Big( \frac 1{\sqrt{\rho }}
\partial _x^2\sqrt{\rho }-(\partial _x\varphi)^2\Big)
+ \Lambda -V(x) -g\rho &= 0,
\\
-i\partial _\tau \rho +\frac{\hbar ^2}m\partial _x
\left( \rho \partial_x\varphi \right) &= 0.
\endaligned
\eqno{(25)}
$$
Let $\rho_0$ and $\varphi_0$ to denote some solutions of the couple
of equations (25). Substituting $\rho_0, \varphi_0$ into the effective
action (24), one obtains
$$
S_{\eff}[\rho _0,\varphi _0]=-\beta F_{nc}(\mu )+\frac g2
\int\limits_0^\beta d\tau\int dx\rho _0^2.
\eqno{(26)}
$$
Here $F_{nc}(\mu )$ is the free energy of the non-ideal gas of
the over-condensate quasi-particles. The total free energy of the system is
$F(\mu )=-\frac{1}{\beta} S_{\eff}[\rho _0,\varphi _0]$ \cite{18}.

Let us use the Thomas--Fermi approximation, which is valid at
sufficiently low temperatures, and drop out the kinetic term
$(\partial _x^2\sqrt{\rho })/\sqrt{\rho }$ in the first equation in (25).
Solution with $\partial _\tau \rho =0=\partial_\tau\varphi$ appears,
provided the velocity field ${\bf v}=m^{-1}\partial _x\varphi$ in (25) is
taken equal to zero. In this case, equations (25) lead to the density of
the condensate described by the solution (14), where the chemical
potential $\mu$ is replaced by $\Lambda$:
$$
\rho _{TF}(x)\equiv\frac{\Lambda}{g}\,
\widetilde \rho _{TF}(x)=\frac \Lambda g \Bigl(1-%
\frac{x^2}{R_c^2}\Bigr) \Theta \Bigl(1-\frac{x^2}{R_c^2}\Bigr).
\eqno{(27)}
$$
Explicit form of the external potential $V(x)=\frac m2\Omega ^2x^2$
is taken into account in expression (27). The form of the solution (27)
means that the quasi-condensate occupies the domain $|x|\le R_c$ at
zero temperature. The length $R_c$ defines the boundary of this domain,
$R_c^2\equiv \frac{2\Lambda }{m\Omega ^2}$ (in three dimensional space,
this would correspond to a spherical distribution of the condensate).
In the homogeneous case given by the limit $1/R_c\to 0$, the Thomas--Fermi
solution $\rho _{TF}(x)$
is transformed into the density $\rho_{TF}(0) =\Lambda/g$, which
coincides with the density of the homogeneous Bose gas \cite{21}.

Following the initial splitting (5), we suppose that
the thermal fluctuations in vicinity of the stationarity point
(27) are small, and therefore an analogous splitting can be written
for the condensate density also:
$$
\rho _0(x,\tau )=\rho _{TF}(x)+\pi _0(x,\tau ).
\eqno{(28)}
$$
The Gross--Pitaevskii equations (25) linearized in a
vicinity of the equilibrium solution $\rho_{TF}=\rho_{TF}(x),
\varphi =\const$, then takes the form:
$$
\aligned
i\partial _\tau \varphi _0-g\pi _0+
\frac{\hbar ^2}{4m\rho _{TF}}\partial_x^2\pi _0 &=0,
\\
i\partial _\tau \pi _0-\frac{\hbar ^2}m\partial _x
(\rho _{TF}\partial_x\varphi _0) &=0.
\endaligned  
\eqno{(29)}
$$
Eliminating $\varphi _0$ and dropping out the terms proportional to
$\hbar ^4$, we go over from (29) to the {\it Stringari thermal equation}
\cite{30}:
$$
\frac 1{\hbar ^2v^2}\partial _\tau ^2\pi _0+\partial _x\Big(\Big( 1-
\frac{x^2}{R_c^2}\Big) \partial _x\pi _0\Big) =0,
\eqno{(30)}
$$
where the parameter $v$ has a meaning of the sound velocity in the
center of the trap:
$$
v^2\equiv\frac{\rho _{TF}(0)g}m=\frac \Lambda m.
\eqno{(31)}
$$

The substitution $\pi_0=e^{i\omega \tau }u(x)$ transforms (30) into
the Legendre equation:
$$
-\frac{\omega ^2}{\hbar ^2v^2}u(x)+\frac d{dx}\Big(\Big(1-\frac{x^2}{%
R_c^2}\Big) \frac d{dx}u(x)\Big) =0.
\eqno{(32)}
$$
Since the Thomas--Fermi solution (27) is different from zero
only at $|x|\le R_c$, we shall consider equation (32) at
$x \in [-R_c, R_c]\subset {\Bbb R}$, as well. After an analytical continuation
$\omega \rightarrow iE$, equation (32) possesses the polynomial solutions,
which are given by the Legendre polynomials $P_n( x/R_c)$, if and only if
$$
\Big( \frac{R_c}{\hbar v}\Big) ^2E^2\equiv \frac 2{\hbar ^2\Omega ^2}
E^2=n(n+1),    \qu n\geq 0.
\eqno{(33)}
$$
In other words, equation (32) leads to the spectrum
of the low lying excitations: $E_n=\hbar \Omega \sqrt{\frac{n(n+1)}2}$,
$n\ge 0$ \cite{31}. Notice that the corresponding equation for the
homogeneous Bose gas is obtained after a formal limit $1/R_c\to 0$
in (32) at finite $x$. Provided the latter is still considered for the
segment $[-R_c, R_c]\ni x$ with a periodic boundary condition for $x$,
we arrive at the discrete spectrum of the following form:
$E_k=\hbar v k$, where $k$ is the wave number, $k=(\pi/R_c)n$, $n\in{\Bbb Z}$.

\head
\S3.~The two-point thermal correlation function
\endhead

Let us go over to the main problem of the present paper ---  to the
calculation of the two-point thermal correlation function of spatially
non-homogeneous Bose gas described by the Hamiltonian (1):
$$
\Gamma (x_1,\tau _1;x_2,\tau _2)\equiv \langle T_\tau
\widehat \psi ^{\dagger}(x_1,\tau _1)\widehat \psi (x_2,\tau _2)\rangle,
\eqno{(34)}
$$
where $T_\tau$ is a ``$\tau$-chronological'' ordering with respect
of the imaginary time $\tau$, and the angular brackets $\langle\,,
\rangle$ correspond to averaging with respect of the Gibbs
distribution \cite{6}. We can express the correlator $\Gamma
(x_1,\tau _1;x_2,\tau _2)$ as a ratio of two functional integrals
\rom{[21--24]}:
$$
\Gamma (x_1,\tau _1;x_2,\tau _2)=\frac{
\int e^{S[\psi ,\bar \psi ]}\bar \psi
(x_1,\tau _1)\psi (x_2,\tau _2){\cal D}\psi {\cal D}\bar\psi}
{\int e^{S[\psi ,\bar \psi ]}{\cal D}\psi {\cal D}\bar\psi},
\eqno{(35)}
$$
where the action $S[\psi ,\bar \psi ]$ is defined in (3).

We are interested in the behaviour of the correlators at the
distances considerably smaller in comparison with the size of the
whole domain occupied by the condensate. The main contribution to
the behaviour of the correlation functions is due to the low lying
excitations at sufficiently low temperatures \cite{21}. To
calculate $\Gamma (x_1,\tau _1;x_2,\tau _2)$ (35), we use the
method of successive functional integration first over the
high-energy excitations $\bar\psi_e$, $\psi_e$, and then over the
low-energy excitations $\bar\psi _o$, $\psi _o$ \cite{21, 24, 25}.
We find that in the leading approximation the correlator we are
interested in looks as follows \cite{21, 25}:
$$
\Gamma (x_1,\tau _1;x_2,\tau _2)\simeq
\frac{\int e^{S_{\eff}[\psi _o,\bar \psi _o]}
\bar \psi _o(x_1,\tau _1)\psi _o(x_2,\tau _2){\cal D}\psi _o{\cal D}
\bar\psi_o}
{\int e^{S_{\eff}[\psi _o,\bar \psi _o]}{\cal D}\psi _o{\cal D}\bar \psi _o},
\eqno{(36)}
$$
where $S_{\eff}[\psi _o,\bar \psi _o]$ is the effective action
(22). We can rewrite (36) in terms of the density--phase variables
(23), and then represent the integrand in the nominator in the
form of a single exponential:
\begin{align*}
&\Gamma (x_1, \tau_1; x_2, \tau_2)
\\
&{\simeq}\frac{\int \exp{\Big(S_{\eff}[\rho,\varphi]{-}
i\varphi (x_1,\tau _1){+}i\varphi (x_2,\tau _2){+}
\frac 12\ln \rho (x_1,\tau_1){+}\frac 12\ln \rho (x_2,\tau _2)\Big)}
{\cal D}\rho {\cal D}\varphi }
{\int \exp{\bigl(S_{\eff}[\rho,\varphi ]\bigr)}{\cal D}\rho {\cal D}\varphi }.
\tag{37}
\end{align*}
Here $S_{\eff}[\rho,\varphi ]$ is the effective action (24).

Since the fluctuations of the density are suppressed at
sufficiently low temperatures \cite{28}, one can replace $\ln \rho
(x_1,\tau _1)$, $\ln \rho (x_2, \tau _2)$ in (37) by $\ln \rho
_{TF}(x_1)$, $\ln \rho _{TF}(x_2)$, where the density $\rho _{TF}$
is defined by (27). In accordance with the variational principle
suggested in \cite{25}, we shall estimate the functional integrals
in (37) by the stationary phase method, and we shall consider only
a leading approximation. For the correlation function $\Gamma
(x_1, \tau_1; x_2, \tau_2)$, we obtain the following estimation:
\begin{align*}
\Gamma (x_1,\tau _1;x_2,\tau _2) & {\simeq}
         \sqrt{\rho _{TF}(x_1)\rho _{TF}(x_2)}
\\[0.2cm]
&\qu{\times}\exp \bigl(-S_{\eff}[\rho _0,\varphi_0]{+}S_{\eff}[\rho_1,\varphi_1]
{-} i\varphi_1(x_1,\tau_1){+}i\varphi_1(x_2,\tau_2)\bigr),
\tag{38}
\end{align*}
where the variables $\rho _0,\varphi _0$ are defined by the extremum
condition $\delta(S_{\eff}[\rho ,\varphi ])=0$, and therefore they
just satisfy the stationary Gross--Pitaevskii equations (25). For
$S_{\eff}[\rho _0,\varphi _0]$ we use the expression (26).

The fields $\rho_1, \varphi_1$ are defined by the extremum
condition:
$$
\delta \Bigl( S_{\eff}[\rho ,\varphi ]-i\varphi (x_1,\tau _1)+i\varphi
(x_2,\tau _2) \Bigr)=0.
\eqno{(39)}
$$
This variational equation leads to another couple of equations of
the Gross--Pitaevskii type. One of these equations turns out to be
a non-homogeneous equation with the $\delta$-like source, while
another one --- a homogeneous equation. In fact, the homogeneous
equation appears due to a requirement of vanishing of the
coefficient at the variation $\delta\rho(x,\tau )$,
$$
 i\partial _\tau \varphi +\frac{\hbar ^2}{2m}\Big( \frac 1{\sqrt{\rho }}%
\partial _x^2\sqrt{\rho }-(\partial _x\varphi) ^2\Big)
+\Lambda -V(x) -g\rho =0.
\eqno{(40)}
$$
In its turn, the non-homogeneous equation is defined by vanishing
of the coefficient at the variation $\delta\varphi (x,\tau )$:
$$
-i\partial _\tau \rho +\frac{\hbar ^2}m\partial _x\left( \rho
\partial _x\varphi \right)  =i\delta (x-x_1)\delta (\tau -\tau _1)-
i\delta (x-x_2)\delta (\tau -\tau_2).
\eqno{(41)}
$$
Substituting the solutions $\rho_1, \varphi_1$, which respect (40), (41),
into the effective action (24), one gets
$$
S_{\eff}[\rho _1,\varphi _1]=-\beta F_{nc}(\mu )-1
+\frac g2\int\limits_0^\beta d\tau\int dx\rho _1^2.
\eqno{(42)}
$$

Further, it can consistently be assumed that the solution $\rho
_1(x,\tau )$ can be represented as a sum of $\rho _{TF}(x)$ and of
a weakly fluctuating part provided the boundary $R_c$ is far from
beginning of coordinates: $\rho _1(x,\tau ) = \rho
_{TF}(x)+ \pi_1(x,\tau )$ (for a comparison, see~(28)). Therefore,
the terms $\sqrt{\pi _1}\partial _x^2\sqrt{\pi _1}$ and $\partial
_x\pi _1 \partial _x\varphi _1$ are small and can be omitted.
Taking into account a linearization near the Thomas--Fermi
solution, one can finally re-write equations (40) and (41) as a
couple of the following equations:
\begin{align*}
i\partial _\tau \varphi _1-g\pi _1-\frac{\hbar ^2}{2m}\left( \partial
_x\varphi _1\right) ^2 &=0,
\tag{43.1}
\\
-i\partial _\tau \pi _1+\frac{\hbar ^2}m\partial _x\left( \rho _{TF}\partial
_x\varphi _1\right) &=i\delta (x-x_1)\delta (\tau -\tau _1)-i\delta
(x-x_2)\delta (\tau -\tau _2).
\tag{43.2}
\end{align*}

Differentiating (43.1) over $\tau $, substituting the result
into (43.2), and dropping out the terms of higher orders in
$g$ and $\hbar ^2$, one obtains the following equation:
$$
\frac 1g\partial _\tau ^2\varphi _1+\frac{\hbar ^2}m\partial _x\left( \rho
_{TF}(x)\partial _x\varphi _1\right) =
i\delta (x-x_1)\delta (\tau -\tau_1)-i\delta (x-x_2)
\delta (\tau -\tau _2).  
\eqno{(44)}
$$
It is convenient to rewrite it as follows:
$$
\frac 1{\hbar ^2v^2}\partial _\tau ^2\varphi _1{+}\partial
_x\left( \widetilde \rho_{TF}(x)\partial_x\varphi_1\right){ = }
\,i\frac{mg}{\hbar ^2\Lambda }\Bigl\{\delta (x-x_1) \delta
(\tau{-}\tau_1){-}\delta (x{-}x_2)\delta (\tau{-}\tau_2)\Bigr\},
\eqno{(45)}
$$
where $v$ means the sound velocity in the center of the trap (31),
and $\widetilde\rho _{TF}$ is defined by (27). More specifically,
solutions of equations (44), (45) depend on the coordinates of the
$\delta $-like sources in Right Hand Side, i.e., on $x_1$,
$\tau_1$, $x_2$, $\tau _2$: $\varphi _1(x,\tau )\equiv \varphi
_1(x,\tau;x_1,\tau _1,x_2,\tau _2)$. Now, with the help of
equations (26), (42) and (43), one can calculate the contribution,
which is a part in the exponent in~(38):
\begin{align*}
&-S_{\eff}[\rho _0,\varphi _0]+S_{\eff}[\rho _1,\varphi _1]
\\
&\qu\simeq \frac g2\int\limits_0^\beta d\tau \int dx(\rho_1^2-\rho_0^2)
 =\frac g2\int\limits_0^\beta d\tau \int dx(\rho _1-\rho _0)
(\rho _1+\rho _0)
\\
&\qu\simeq g\int\limits_0^\beta d\tau \int dx\pi _1\rho _0
=\int\limits_0^\beta d\tau \int
dx\Big(i\partial _\tau \varphi _1-\frac{\hbar ^2}{2m}(
\partial_x\varphi _1) ^2\Big) \rho _0
\\
&\qu=-\frac{\hbar ^2}{2m}\int\limits_0^\beta d\tau \int dx\rho _0\left(
\partial_x\varphi _1\right) ^2=\frac{\hbar ^2}{2m}\int\limits_0^\beta d\tau
\int dx\varphi_1(x)\partial_x\bigl( \rho_0(x)\partial_x\varphi_1(x)\bigr)
\\
&\qu=\frac i2\bigl( \varphi_1(x_1,\tau_1)-\varphi_1(x_2,\tau_2)\bigr) .
\tag{46}
\end{align*}
Substituting (46) into (38), one obtains the following approximate formula for
the correlator:
$$
\Gamma (x_1, \tau_1; x_2, \tau_2) \simeq \sqrt{\rho_{TF}(x_1)\rho_{TF}(x_2)}
\,\exp{\Bigl(-\frac i2\bigl( \varphi_1(x_1,\tau_1)-\varphi_1(x_2, \tau_2)
\bigr)\Bigr)}.
\eqno{(47)}
$$

It is natural to represent the solutions of equations (44), (45)
in terms of the solution $G(x, \tau ; x^{\prime}, \tau^{\prime})$
of the equation
$$
\frac 1{\hbar ^2v^2}\partial _\tau ^2G(x,\tau ;x^{\prime },\tau
^{\prime }){+}\partial _x\Big(\Bigl( 1{-}\frac{x^2}{R_c^2}\Bigr) \partial
_xG(x,\tau;x^{\prime },\tau ^{\prime })\Big){=}\frac g{\hbar ^2v^2}\delta
(x{-}x^{\prime })\delta (\tau{-}\tau ^{\prime }).
\eqno{(48)}
$$
Bearing in mind the homogeneous equation (30), we shall call (48) as
{\it non-homogeneous Stringari equation}. As a result, the
representation (47) can be re-written as follows:
\begin{align*}
&\Gamma (x_1,\tau _1;x_2,\tau _2)
\\
&\qu\simeq \sqrt{\rho_{TF}(x_1)\rho_{TF}(x_2)}
\exp\Bigl(-\frac 12\big(G(x_1,\tau _1;x_2,\tau _2)
+G(x_2,\tau_2;x_1,\tau _1)\big)
\\
&\hphantom{\simeq \sqrt{\rho_{TF}(x_1)\rho_{TF}(x_2)}\exp\exp}
+\frac 12G(x_1,\tau _1;x_1,\tau _1)
+ \frac 12 G(x_2,\tau _2;x_2,\tau _2)\Bigr).
\tag{49}
\end{align*}
As it is clear after \cite{8}, the function $G(x_1,\tau _1;x_2,\tau _2)$
has a meaning of the correlation function of the phases:
$$
G(x_1,\tau _1;x_2,\tau_2)=- \bigl\langle
\varphi (x_1,\tau _1) \varphi (x_2,\tau _2)\bigr\rangle,
\eqno{(50)}
$$
where the angle brackets in Right Hand Side should be understood
as an averaging with respect to the weighted measure ${\cal D}\rho {\cal
D}\varphi$ $\exp{\bigl(S_{\eff}[\rho,\varphi ]\bigr)}$. Substituting (50)
into (49), we obtain the known approximate formula for the correlator
(34) \cite{8, 21}:
$$
\big\langle T_\tau\widehat \psi^{\dagger}(x_1, \tau_1) \widehat
\psi (x_2, \tau_2)\big\rangle \,\simeq\, \sqrt{\rho_{TF}(x_1)
\rho_{TF}(x_2)}\,\exp{\Bigl(-\frac 12 \bigl\langle \left( \varphi
(x_1,\tau_1)-\varphi (x_2,\tau_2) \right)^2 \bigr\rangle\Bigr)}.
\eqno{(51)}
$$

Notice that the terms in the exponent in (49) have different
meanings. The terms $G(x_1, \tau_1; x_2, \tau_2)$ and $G(x_2,
\tau_2; x_1,\tau_1)$ depend on the differences of the arguments of
two-point correlation function and therefore are responsible for
the behaviour of the correlator at large distances. The terms
$G(x_1, \tau_1; x_1, \tau_1)$, $G(x_2, \tau_2; x_2, \tau_2)$ each
depend only on a single set of the coordinates and thus
contribute to the amplitudes only. The Green function $G(x, \tau ;
x^{\prime }, \tau ^{\prime })$ depends, in fact, on the difference
of $\tau$ and $\tau^{\prime}$ due to the invariance under
the shifts of $\tau$ (see. (48)). Therefore, only a dependence
on the spatial coordinates remains provided the corresponding
thermal arguments coincide. Then, it is possible to represent the
correlation function as follows:
$$
\Gamma (x_1, \tau_1; x_2, \tau_2) \,\simeq\,\sqrt{\widetilde \rho
(x_1) \widetilde\rho (x_2)} \,\exp{\Bigl(-\frac 12\left( G(x_1,
\tau_1; x_2, \tau_2) +
 G(x_2, \tau_2; x_1, \tau_1) \right) \Bigr) },
\eqno{(52)}
$$
where $\widetilde \rho (x_1)$, $\widetilde \rho (x_2)$ are the
renormalized densities \cite{18}. The solution $G( x_1, \tau_1;
x_2, \tau_2)$ of equation (48) is defined up to a purely imaginary
additive constant, which has a meaning of a global phase. As it is
seen from (51), this constant does not influence the fluctuations.

\head
\S4.~The asymptotics of the correlation function
\endhead

Therefore, the problem concerning the study of the asymptotical
behaviour of the two-point thermal correlation function
$\Gamma (x_1,\tau _1; x_2,\tau _2)$ (34), which is given by the
representation (37), is reduced to solution of the
non-homogeneous Stringari equation (48). The corresponding
answer (or its asymptotics) should be subsequently substituted
into (52). In the present section, we shall obtain explicitly
solutions of equation (48), and we shall consider the corresponding
representations for the asymptotics of $\Gamma (x_1,\tau _1;x_2,\tau _2)$.
Let us begin with the limiting case of
equation (48), which corresponds to a homogeneous Bose gas.

\subhead{1.~The homogeneous Bose gas}
\endsubhead
In this subsection we shall consider the asymptotical behaviour of
the correlation function of the homogeneous Bose gas. As it was
mentioned above, the homogeneous case corresponds to $V(x)\equiv
0$, and the related equation appears from (48) at $1/R_c\to 0$:
$$
\frac 1{\hbar ^2v^2}\partial _\tau ^2G(x,\tau ;x^{\prime },\tau ^{\prime
})+\partial _x^2G(x,\tau ;x^{\prime },\tau ^{\prime })=\frac g{\hbar ^2v^2}
\delta (x-x^{\prime })\delta (\tau -\tau ^{\prime }).
\eqno{(53)}
$$
We consider (53) for the domain $[-R_c, R_c]\times [0, \beta]$
$\ni$ $(x, \tau)$ with the periodic boundary conditions for each
variable (in other words, we consider (53) on the torus
$S^1\times S^1$ $\ni$ $(x, \tau)$). The $\delta$-functions in
Right Hand Side of (53) are treated as the corresponding periodic
$\delta$-functions. This allows us to represent the solution of
this equation as the formal double Fourier series:
\begin{align*}
&G(x, \tau ; x^{\prime }, \tau^{\prime})  =
\Bigl(\frac{-g}{\hbar^2 v^2}\Bigr)\, (2 \beta R_c)^{-1}
\sum\limits_{\omega, k}
  \frac{e^{i\omega(\tau - \tau^{\prime})+ik(x-x^{\prime})}}
   {\omega^2/(\hbar^2 v^2)+k^2} \\
&\qu= \Bigl(\frac{-g}{2 \beta R_c}\Bigr)\,\sum\limits_{\omega, k}
 \frac{e^{i\omega(\tau-\tau^{\prime})+ik(x-x^{\prime})}}
 {\omega^2+E_k^2},
\tag{54}
\end{align*}
where $\omega=(2\pi/\beta) l$, $l\in {\Bbb Z}$. The notation for the
energy $E_k=\hbar v k$, where $k=(\pi/R_c) n$, $n\in {\Bbb Z}$, is used
in the representation (54). Besides, the representation (54)
requires a regularization, which consists in neglection of the
term given by $\omega = k = 0$.

Using (54), one can deduce two important asymptotical
representations for the Green function. In the limit of zero
temperature and of infinite size of the domain occupied by the
Bose gas, one can go over to the asymptotics of
$\Gamma (x_1,\tau_1; x_2,\tau _2)$. When a strong inequality
$\beta^{-1}\equiv k_BT\gg \hbar v/R_c$ is valid, we obtain:
\begin{align*}
&G(x, \tau; x^{\prime}, \tau^{\prime})
\\
&\qu\simeq \frac g{2\pi \hbar v}\ln
\Big\{2\Big| \sinh \frac \pi {\hbar \beta v}(| x-x^{\prime }|
+i\hbar v(\tau -\tau ^{\prime })) \Big|
\Big\} -\frac g{4\beta R_c}\,\frac{|x-x^{\prime }|^2}{\hbar^2v^2}+{\cal C},
\tag{55}
\end{align*}
where $|x-x^{\prime }|\leq 2 R_c,\,|\tau -\tau ^{\prime }|\leq \beta$, and
${\cal C}$ is some constant, which is not written explicitly.
When an opposite inequality $\beta^{-1}\equiv k_BT\ll \hbar v/R_c$ is valid,
we obtain:
\begin{align*}
&G(x, \tau ; x^{\prime}, \tau^{\prime})
\\
&\qu\simeq \frac g{2\pi \hbar v}\ln
\Big\{2\Big| \sinh \frac{i\pi}{2 R_c}(| x-x^{\prime }|
+i\hbar v(\tau -\tau ^{\prime })) \Big| \Big\}
-\frac g{4 \beta R_c}\,|\tau-\tau^{\prime }|^2+{\cal C}^{\prime}\,,
\tag{56}
\end{align*}
where $|x-x^{\prime }|\leq 2 R_c,\,|\tau -\tau ^{\prime }|\leq \beta$,
and ${\cal C}^\prime$ is another constant.

Let us substitute the estimate (55) into the representation (52)
and take simultaneously the limit $\beta\hbar v/R_c$ $\to 0$
(the size is growing faster than inverse temperature). Then, we
obtain the following expression for the correlator in question:
$$
\Gamma (x_1,\tau _1;x_2,\tau _2)\simeq
\sqrt{\widetilde\rho(x_1) \widetilde\rho(x_2)}\,\,
{\Big| \sinh \frac \pi {\hbar \beta v}\left(|x_1-x_2|
+i\hbar v(\tau_1 -\tau_2)\right) \Big|}^{-g/2\pi \hbar v}.
\eqno{(57)}
$$
Further, applying the relation (56) and taking the limit
$R_c/(\beta\hbar v) \to 0$ (the inverse temperature grows faster
than the size), we obtain for $\Gamma (x_1, \tau_1; x_2,
\tau_2)$:
$$
\Gamma (x_1, \tau_1; x_2, \tau_2)\simeq
\sqrt{\widetilde\rho(x_1) \widetilde\rho(x_2)}\,\,
{\Big| \sinh \frac{i\pi}{2 R_c}\left(|x_1-x_2|
+i\hbar v(\tau_1 -\tau_2)\right) \Big|}^{-g/2\pi \hbar v}.
\eqno{(58)}
$$

It follows from (57) and (58), that in the limit of zero
temperature, $(\hbar\beta v)^{-1}\to 0$, and of infinite
size, $1/R_c\to 0$, the two-point correlation function behaves like
$$
\Gamma (x_1,\tau _1;x_2,\tau _2)\simeq
\frac{\sqrt{\widetilde\rho(x_1) \widetilde\rho(x_2)}}
{||x_1-x_2|+i\hbar v(\tau_1 -\tau_2)|^{1/\theta}}.
\eqno{(59)}
$$
The latter formula is valid in the limit $\beta\hbar v/R_c\to 0$,
as well as in the limit $R_c/(\beta\hbar v)\to 0$. In (59),
$\theta$ denotes the critical exponent: $\theta\equiv 2\pi\hbar
v/g$, and the arguments $x_1$ and $x_2$, $\tau_1$ and $\tau_2$,
are assumed to be sufficiently close each to other. Using the
notations $v={\sqrt{\Lambda/m}}$ for the sound velocity and
$\rho=\Lambda/g$ for the density of the homogeneous ideal Bose
gas, we obtain for the critical exponent the following universal
expression \cite{11, 12}:
$$
\theta=\frac{2\pi\hbar \rho}{m v}\,.
\eqno{(60)}
$$

\subhead{2.~The trapped Bose gas. High temperature case:
$k_BT\gg \hbar v/R_c$}
\endsubhead
Let us consider the case of non-homogeneous Bose gas, which is
described by the Hamiltonian (1) with the external potential
$V(x)\equiv \frac m2 \Omega ^2x^2$. In the present subsection we
are mainly following the content of the paper \cite{20}. Let us
consider the non-homogeneous Stringari equation (48) and write it
again, for convenience:
$$
\frac 1{\hbar ^2v^2}\partial _\tau ^2G(x,\tau ;x^{\prime },\tau ^{\prime
})+\partial _x\Big(\Bigl( 1-\frac{x^2}{R_c^2}\Bigr) \partial _xG(x,\tau
;x^{\prime },\tau ^{\prime })\Big) =\frac g{\hbar ^2v^2}\delta
(x-x^{\prime })\delta (\tau -\tau ^{\prime }).
\eqno{(61)}
$$
We consider (61) for the arguments $(x, \tau)$ $\in$ $[-R_c,
R_c]\times [0, \beta]$ with the periodic boundary condition only
with respect to $\tau$ (contrary to equation (53), $\delta
(x-x^\prime)$~ is a usual Dirac's $\delta$-function with a support
at the point $x^\prime\in {\Bbb R}$). The Green function
satisfying (61) can be written as a formal Fourier series:
$$
G(x,\tau ;x^{\prime },\tau ^{\prime })=\frac 1\beta \sum_\omega e^{i\omega
(\tau -\tau ^{\prime })}G_\omega (x,x^{\prime }),
\eqno{(62)}
$$
where $\omega =(2\pi/\beta) l$, $l \in {\Bbb Z}$. The spectral density
$G_\omega (x,x^{\prime })$ in (62) is then governed by the equation
$$
-\frac{\omega ^2}{\hbar ^2v^2}G_\omega (x,x^{\prime })
+ \frac{d}{dx}\Big(\Bigl(1-\frac{x^2}{R_c^2}\Bigr)
\frac{d}{dx}G_\omega (x,x^{\prime })\Big) = \frac{g}{\hbar^2 v^2}
\delta (x-x^{\prime }).
\eqno{(63)}
$$

Solution of equation (63) can be obtained in terms of the Legendre
functions of the first and second kind, $P_\nu (x/R_c)$ and $Q_\nu
(x/R_c)$, which are linearly independent solutions of the
homogeneous Legendre equation (32). As a result, we get:
$$
G_\omega (x,x^{\prime })
=\frac{gR_c}{2\hbar^2 v^2}\,\epsilon (x-x^{\prime})
\Big\{ Q_\nu \Bigl(\frac x{R_c}\Bigr)
P_\nu\Bigl(\frac{x^{\prime }}{R_c}\Bigr)-Q_\nu
                           \Bigl(\frac{x^{\prime }}{R_c}\Bigr)
P_\nu\Bigl(\frac x{R_c}\Bigr)\Big\},
\eqno{(64)}
$$
where $\nu$ looks as follows:
$$
\nu =-\frac 12+
\sqrt{\frac 14-\Bigl( \frac{R_c}{\hbar v}\Bigr) ^2\omega ^2},
$$
and $\epsilon (x-x^{\prime })$ is the sign function
$\epsilon (x)\equiv \sign(x)$. Validity of the solution (64) can
be verified by direct substitution of (64) into (63), where the
following expression for the Wronskian of two linearly independent
solutions $P_\nu$ and $Q_\nu$ \cite{32} should be used:
$$
P_\nu (y) \frac{d}{dy}\,Q_\nu (y) -
        Q_\nu (y) \frac{d}{dy}\,P_\nu (y)\,=\, (1-y^2)^{-1},
$$
and the rule of differentiation of the sign function is:
$(d/dx) \,\epsilon (x)=2\delta (x)$.

Provided a dependence on $\tau$ is neglected, an equation, which
arises as a result of calculation of the correlation
function accordingly to \cite{25}, looks analogously to equation (61),
but a factor $\beta^{-1}$ is present in its Right Hand Side instead of
$\delta (\tau-\tau^\prime)$. In this case, the corresponding solution
of the non-homogeneous equation (i.~e., the fundamental solution)
$G(x; x^{\prime})$ takes the form
$$
G(x; x^{\prime })=\frac 1\beta G_0(x,x^{\prime }),
\eqno{(65.1)}
$$
where
\begin{align*}
G_0(x,x^{\prime }) &=\frac{gR_c}{2 \hbar^2 v^2}\,\epsilon (x-x^{\prime})
\Big\{ Q_0\Bigl(\frac x{R_c}\Bigr)-
        Q_0\Bigl(\frac{x^{\prime }}{R_c}\Bigr)\Big\}
\\[0.3cm]
&=\frac{gR_c}{(2\hbar v)^2}\ln \left[\frac{\Bigl( 1+
\frac{|x-x^\prime|}{2R_c}\Bigr) ^2-\frac{(x+x^\prime)^2}{4R_c^2}}
{\Big(1-\frac{|x-x^\prime|}{2R_c}\Big)^2-
\frac{(x+x^\prime)^2}{4R_c^2}}\right].
\tag{65.2}
\end{align*}
An explicit form for the simplest Legendre functions $P_0(x)$ $=$
$1$ and $Q_0(x)$ $=$ $\frac 12\ln \frac{1+x}{1-x}$ is essential
for obtaining the relations (65). The fundamental solution $G(x;
x^{\prime })$ (65) becomes equal to zero when its arguments
coincide. A substitution of (65) into the representation (49)
gives the following result for the stationary correlation function
\cite{17, 18, 33, 34}:
$$
\Gamma (x_1; x_2) \simeq \sqrt{\rho_{TF}(x_1) \rho_{TF}(x_2)}
\,\, \exp \Biggl(-\frac{g R_c}{\beta (2 \hbar v)^2}
\ln\left[ \frac{1+\frac{|x_1-x_2|}{R_c}-\frac{x_1 x_2}{R_c^2}}
{1-\frac{|x_1-x_2|}{R_c} -\frac{x_1 x_2}{R_c^2}}\right]\Biggr).
\eqno{(66)}
$$

Before studying the behaviour of the correlation function, which depends
on $\tau$, it should be noticed that solutions of the homogeneous
Legendre equation (32) can be added to the Green function (62),
since the latter respects the non-homogeneous equation. In order to
ensure a correct asymptotical behaviour of the spectral density we
are interested in at large $|\omega|$, we are free to add such a term
at $|\omega |\ne 0$ and obtain the following expression:
\begin{align*}
G_\omega (x,x^{\prime }) &=
\frac{gR_c}{2 \hbar^2 v^2} \epsilon (x-x^{\prime })
\Big\{Q_\nu \Bigl(\frac x{R_c}\Bigr) P_\nu \Bigl(\frac{x^{\prime }}{R_c}\Bigr)
-Q_\nu \Bigl(\frac{x^{\prime }}{R_c}\Bigr) P_\nu \Bigl(\frac x{R_c}\Bigr)
\Big\}
\\[0.3cm]
&\qu- i \frac{gR_c}{2 \hbar^2 v^2}
\Big\{\frac 2\pi Q_\nu\Bigl(\frac x{R_c}\Bigr) Q_\nu \Bigl(\frac{x^{\prime }}{R_c}
\Bigr)+\frac \pi 2P_\nu \Bigl(\frac{x^{\prime }}{R_c}\Bigr)P_\nu
\Bigl(\frac x{R_c}\Bigr)\Big\}.
\tag{67.1}
\end{align*}
The Green function given by (62) and (67.1) can be represented in the
form, which allows to study the corresponding asymptotical behaviour.
In the case of strong inequality $\beta^{-1}$ $=$ $k_BT\gg \hbar v/R_c$,
we approximately obtain for non-zero frequencies: $|\omega |\gg \hbar
v/(2R_c)$. Let us take into account the following asymptotics of the
Legendre functions \cite{32, 36}:
\begin{align*}
&\left\{\begin{matrix}
       P_\nu \\ Q_\nu
\end{matrix}   \right\}
(\cos\theta)=\frac{\Gamma (\nu+1)}{\Gamma (\nu+3/2)}\,
\Bigl(\frac{2}{\pi}\Bigr)^{\delta/2}\,\frac{1}{\sin^{1/2}\theta}
\left\{\begin{matrix}
           \sin \\ \cos
\end{matrix}  \right\}
\Big[\Big(\nu +\frac 12\Big)\theta +\frac \pi 4\Big]+O(\nu^{-1})
\\[0.3cm]
&\qq \approx\Bigl(\frac{2}{\pi}\Bigr)^{\delta/2}\,
\frac{1}{\bigl(\nu \sin\theta\bigr)^{1/2}}
\left\{\begin{matrix}
           \sin \\ \cos
\end{matrix} \right\}
\Big[\Big(\nu +\frac 12\Big)\theta +\frac \pi 4\Big],
\tag{67.2}
\end{align*}
where $\varepsilon < \theta < \pi-\varepsilon$, $\varepsilon
> 0$, $|\arg \nu |< \pi/2$, and the notation $\cos \theta \equiv
x/R_c$ is adopted. Only the up or the down expressions must be
simultaneously chosen inside the curly brackets $\{ \dots
\}$ in (67.2). Here $\delta=1$ corresponds to $P_\nu$, and
$\delta=-1$ corresponds to $Q_\nu$. Substituting (67.2) into
(67.1), we determine the behaviour of $G_\omega (x,x^{\prime })$
at large $|\omega |$:
$$
G_\omega (x,x^{\prime })\simeq -\frac g{2\hbar v|\omega |}\frac 1{\sqrt{\sin
\theta \sin \theta ^{\prime }}}\exp \Bigl( -\frac{R_c}{\hbar v}|\omega
|\,|\theta -\theta ^{\prime }|\Bigr).
\eqno{(68)}
$$

When the coordinates $x_1$, $x_2$ are chosen to be far from the
boundary of the trap, $x_1$, $x_2\ll R_c$, but at the same time
the inequalities $|x_1-x_2|\ll \frac{x_1+x_2}2$ and $|x_1-x_2|\ll R_c$
are valid, the corresponding limit should be called as
{\it quasi-homogeneous}. In this case, the function $G_0(x,x^{\prime })$
(65.2) can be approximated up to a second order (including the
latter) as follows:
$$
G_0(x,x^{\prime })\simeq \frac \Lambda {2\hbar ^2v^2\rho _{TF}(S)}
|x-x^{\prime }|.
\eqno{(69)}
$$
Here $S$ means a half-sum of the spatial arguments of the
correlator, $S\equiv \frac{x_1+x_2}2$, and $v$ is the sound
velocity (31). In the quasi-homogeneous limit, equation
(68) can be re-written in the following form:
$$
G_\omega (x,x^{\prime })\simeq -\frac \Lambda {2\hbar v\rho _{TF}(S)}
\frac{\exp{(-(\hbar v)^{-1} |\omega||x-x^{\prime }|)}} {|\omega|}.
\eqno{(70)}
$$
Substitution of (69) and (70) into the series (62) leads to the
answer for the Green function (i.e., for the correlator of the
phases) (50):
$$
G(x,\tau ;x^{\prime },\tau ^{\prime })\simeq\frac \Lambda {2\pi \hbar
v\rho_{TF}(S)}\ln \Big\{ 2\Big| \sinh \frac \pi {\hbar \beta v}
(|x-x^{\prime }| +i\hbar v(\tau -\tau ^{\prime })) \Big|\Big\}.
\eqno{(71)}
$$

Therefore, the Green function (52) takes the following form at
$\beta^{-1} \gg \hbar v/R_c$:
$$
\Gamma (x_1,\tau _1;x_2,\tau _2)\simeq
\frac{\sqrt{\widetilde\rho (x_1) \widetilde\rho (x_2)}}
{| \sinh \frac \pi {\hbar \beta v}(|x_1-x_2|
+i\hbar v(\tau _1-\tau _2))| ^{1/\theta (S)}},
\eqno{(72)}
$$
where the critical exponent $\theta (S)$ depends now only on the
half-sum of the coordinates $S$:
$$
\theta (S)=\frac{2\pi \hbar \rho _{TF}(S)}{mv}.
\eqno{(73)}
$$
The result (72), which is valid for the spatially non-homogeneous case, is in
a correspondence with the estimation (57) obtained above for the
homogeneous Bose gas. Therefore, the estimation (72) is also concerned with
validity of the appropriate condition that the size of the domain occupied
by the Bose condensate grows faster than inverse temperature, i.e.,
with the condition $\hbar \beta v/R_c \to 0$.

The relation (72) can be simplified for two important limiting
cases. Provided the condition
$$
1\ll \frac{| x_1-x_2|}{\hbar \beta v} \ll\frac{R_c}{\hbar \beta v}
\eqno{(74)}
$$
is fulfilled in the quasi-homogeneous case, we obtain from (72) that the
correlation function decays exponentially:
$$
\aligned
\Gamma (x_1,\tau _1;x_2,\tau _2) &\simeq
\sqrt{\widetilde\rho (x_1) \widetilde\rho (x_2)}\,\,
\exp \Bigl(-\frac{1}{\xi(S)}
\bigl||x_1-x_2| +i\hbar v(\tau _1-\tau _2)\bigr| \Bigr),
\\
\xi^{-1}(S) &= \frac{\Lambda}{2\beta \hbar ^2v^2\rho_{TF}(S)}.
\endaligned
\eqno{(75)}
$$
The correlation length $\xi(S)$ is defined by the relation (75),
which depends now on the half-sum of the coordinates:
$$
\xi (S)\equiv \frac{\hbar \beta v}\pi \theta (S) =
\frac{2\hbar ^2\beta \rho _{TF}(S)}m.
\eqno{(76)}
$$

In an opposite limit,
$$
\frac{| x_1-x_2|}{\hbar \beta v},
\,\frac{| \tau_1-\tau_2|}{\beta}\ll 1\ll
\frac{R_c}{\hbar \beta v},
\eqno{(77)}
$$
the asymptotics of $\Gamma (x_1,\tau _1;x_2,\tau _2)$
takes the following form:
$$
\Gamma (x_1,\tau _1;x_2,\tau _2)\simeq
\frac{\sqrt{\widetilde\rho (x_1) \widetilde\rho (x_2)}}
{|| x_1-x_2| +i\hbar v(\tau _1-\tau _2)|^{1/\theta (S)}}.
\eqno{(78)}
$$
The obtained asymptotics (78) is analogous to the estimation (59),
which characterizes the spatially homogeneous Bose gas. But the
critical exponent $\theta (S)$ (73) differs from $\theta$ (60),
since it depends on the spatial coordinates.

\subhead{3.~The trapped Bose gas. Low temperature case: $k_BT\ll \hbar v/R_c$}
\endsubhead
Let us pass to another case which also admits investigation of the
asymptotical behaviour of the two-point correlator $\Gamma (x_1,
\tau_1; x_2, \tau_2)$. As in the previous subsection, we begin with
the non-homogeneous Stringari equation (48), which can be written in the
following form:
\begin{align*}
&\partial _\tau ^2 G(x,\tau ;x^{\prime },\tau ^{\prime})+
\frac{1}{\alpha^2}\,\partial_{(x/R_c)} \Big(\Bigl(
1-\frac{x^2}{R_c^2}\Bigr) \partial_{(x/R_c)} G(x,\tau;x^{\prime
},\tau ^{\prime })\Big)
\\
&\qq= \frac g{R_c}\delta
\Bigl(\frac{x-x^{\prime }}{R_c}\Bigr)\delta (\tau -\tau ^{\prime }),
\tag{79}
\end{align*}
where the notation $\alpha\equiv R_c/(\hbar v)$ is introduced.
The asymptotical behaviour can be investigated in two cases, and
these cases may be characterized in terms of $\alpha$: $\beta/
\alpha$ $\ll 1$ (the previous subsection) and $\beta/\alpha$ $\gg
1$ (see below). The functions
$$
\sqrt{n+\frac12}\,P_n\Bigl(\frac{x}{R_c}\Bigr),
\quad n\ge 0,
$$
where $P_n(x/R_c)$ are the Legendre polynomials, constitute a
complete orthonormal system in the space $L_2\,[-R_c, R_c]$.
This fact allows to obtain the following representation for the Green
function $G(x, \tau ; x^{\prime}, \tau^{\prime})$ in the form of
the generalized double Fourier series:
$$
G(x,\tau ;x^{\prime },\tau ^{\prime})=
\Bigl(\frac{-g}{\beta R_c}\Bigr)\,
\sum\limits_\omega \sum\limits_{n=0}^{\infty}
\frac{n+1/2}{\omega^2+E_n^2}\,P_n\Bigl(\frac{x}{R_c}\Bigr)
P_n\Bigl(\frac{x^{\prime}}{R_c}\Bigr)\,e^{i\omega(\tau-\tau^{\prime})}.
\eqno{(80)}
$$
In (80), as well as in (54), (62), the notation
$\sum\limits_\omega$ denotes a sum over the Bose frequencies
$\omega=(2\pi/\beta) l$, $l\in {\Bbb Z}$, and the following
notation for the energy levels (33) is adopted:
$$
E_n=\hbar\Omega{\sqrt{\frac{n(n+1)}{2}}} =
\frac{\sqrt{n(n+1)}}{\alpha}.
\eqno{(81)}
$$
Let us note that a transition from (54) to (80) takes place
provided the wave functions and the dispertion relation are
appropriately substituted.

After summation over the frequencies and after regularization
consisting in neglection of the term corresponding to zero values
of $\omega$ and $n$, $G(x,\tau ;x^{\prime }, \tau ^{\prime})$ (80)
takes the form:
\begin{align*}
&G(x,\tau ;x^{\prime },\tau ^{\prime})
\\
&\,\, = \Bigl(\frac{-g}{\beta R_c}\Bigr)\,
\Big[\Bigl(\frac{\beta}{2\pi}\Bigr)^2\,\sum\limits_{l=1}^\infty
\,\frac{\cos\bigl(\frac{2\pi\Delta\tau}{\beta}\,l\bigr)}{l^2}
+\frac{\beta}2\,\sum\limits_{n=1}^\infty\,
\frac{n+1/2}{E_n}\,P_n\Bigl(\frac{x}{R_c}\Bigr)
P_n\Bigl(\frac{x^{\prime}}{R_c}\Bigr)
\\
&\hphantom{=\Bigl(\frac{-g}{\beta R_c}\Bigr)=\Bigl(\frac{-g}{\beta R_c}\Bigr) \Bigl(\frac{-g}{\beta R_c}\Bigr)}
\times\Big(\coth\Bigl(\frac{\beta}2E_n\Bigr)
\cosh(E_n \Delta\tau)-\sinh(E_n \Delta\tau)\Big)\Big],
\tag{82}
\end{align*}
where $\Delta\tau\equiv |\tau-\tau^{\prime}|$. The obtained
representation (82) admits an investigation for two limiting
cases: $\beta/\alpha \ll 1$ (this case agrees with the estimation
(75) of the previous subsection) and $\beta/\alpha\gg 1$.

Indeed, putting $\tau=\tau^\prime$ in the case $\beta/\alpha \ll
1$, we obtain \cite{35} the following relation:
\begin{align*}
&G(x,\tau ;x^{\prime },\tau)=-\frac{g\beta}{24 R_c}-
\frac{g}{\beta R_c}\sum\limits_{n=1}^\infty\,\frac{n+1/2}{E_n^2}\,
P_n\Bigl(\frac{x}{R_c}\Bigr) P_n\Bigl(\frac{x^{\prime}}{R_c}\Bigr)
\\
&\qu=\frac{g R_c}{2 \beta\hbar^2 v^2} \ln\Big(1+ \frac{|x-x^\prime|}{R_c}-
\frac{x x^\prime}{R_c^2}\Big) +
\frac{g R_c}{\beta\hbar^2 v^2}\Big(\frac12-\ln 2-
\frac{\beta^2}{24 \alpha^2}\Big).
\tag{83}
\end{align*}
Let us observe that Right Hand Side of (83) respects the
non-homogeneous equation
$$
\frac{1}{\alpha^2}\,\partial_{(x/R_c)}
\Big(\Bigl( 1-\frac{x^2}{R_c^2}\Bigr) \partial_{(x/R_c)}
G(x,\tau; x^{\prime },\tau)\Big) = \frac g{\beta R_c}
\Big(\delta\Bigl(\frac{x-x^{\prime }}{R_c}\Bigr)-\frac12\Big).
\eqno{(84.1)}
$$
Provided the spatial arguments in the relation (83) are equated,
one can obtain the following equality:
\begin{align*}
&\frac{\,G(x, \tau; x, \tau) + G(x^\prime, \tau; x^\prime,\tau)\,}{2}
\\[0.3cm]
&\qu=\frac{g \alpha^2}{4\beta R_c}\,\ln\,\Big[\Bigl(
1-\frac{x^2}{R_c^2}\Bigr) \Bigl(
1-\frac{(x^\prime)^2}{R_c^2}\Bigr)\Big] + \frac{g
R_c}{\beta\hbar^2 v^2}\Big(\frac12-\ln 2- \frac{\beta^2}{24
\alpha^2}\Big). \tag{84.2}
\end{align*}
Direct substitution demonstrates that (84.2) satisfies an equation, where
Left Hand Side is the same as in (84.1), while only the constant term
$\frac{-g}{2\beta R_c}$ is present in Right Hand Side.
Therefore, a subtraction of Right Hand Side of (84.2) from Right
Hand Side of (83) results exactly  in the fundamental solution
$G(x; x^\prime)$ (65). The Green function $G (x; x^\prime)$
becomes equal to zero at $x=x^\prime$, and it respects an
equation of the type of (84.1) but with $\frac{g}{\beta} \delta
(x-x^\prime)$ in Right Hand Side.

With the help of the series (83), we may bring the exponent in (49)
into the form:
$$
\Bigl(\frac{-g}{2 \beta R_c}\Bigr)\,
\sum\limits_{n=1}^\infty\,\frac{n+1/2}{E_n^2}\,
\Big(P_n\Bigl(\frac{x}{R_c}\Bigr) -
P_n\Bigl(\frac{x^{\prime}}{R_c}\Bigr)\Big)^2.
\eqno{(85)}
$$
As it is seen from (83) and (84.2), the relation (85) is nothing
else but the fundamental solution $G(x; x^\prime)$ (65) taken with
an opposite sign. Thus, the representation (83), being substituted
into (49), leads to the formula for the ``equal-time'' correlator
$\Gamma (x_1, \tau; x_2, \tau)$, which has the same form as
Right Hand Side of (66). Taking into account the quasi-homogeneity
conditions, which were discussed in the previous subsection, and
recalling the corresponding estimation (69), we obtain for $G(x;
x^\prime)$:
$$
G (x,x^{\prime })\simeq \frac \Lambda {2\beta\hbar^2v^2 \rho _{TF}(S)}
|x-x^{\prime }|.
$$
In its turn, the latter formula leads to the estimation:
$$
\Gamma (x_1, \tau; x_2, \tau)\simeq
\sqrt{\rho_{TF}(x_1) \rho_{TF}(x_2)}\,\,
\exp{\Bigl(-\frac{1}{\xi(S)} |x_1-x_2|\Bigr)},
\eqno{(86)}
$$
where $\xi(S)$ is the correlation length (76). As a result, equations (75)
and (86) demonstrate an agreement of the estimations based on two
different representations for the Green function $G(x,\tau ;x^{\prime },
\tau^\prime)$: the first one is in the form of the trigonometric Fourier
series (62) (where either (64) or (67) is used to express
$G_\omega (x, x^{\prime })$), and the second one is in the form of the
series (82), which runs over the principle quantum numbers (obtained from
the generalized double Fourier series (80)).

Now let us turn to the case $\beta/\alpha\gg 1$, where $\beta
E_n\gg 1$, $\forall n=1, 2, \dots$  In other words, let us suppose that
$k_B T\ll E_n$ and, so, $k_B T\ll \hbar\Omega$. Then, from (82) one obtains:
\begin{align*}
&G(x,\tau ;x^{\prime },\tau^\prime) \,=\,\frac{-g \beta}{4 R_c}
\Big[\Bigl(\frac12-\frac{\Delta\tau}{\beta}\Bigr)^2-\frac1{12}\Big]
\\
&\qq-\frac{g}{2\hbar v} \sum\limits_{n=1}^\infty\,\frac{n+1/2}{\sqrt{n(n+1)}}\,
P_n\Bigl(\frac{x}{R_c}\Bigr) P_n\Bigl(\frac{x^{\prime}}{R_c}\Bigr)\,
\exp\Bigl(-\sqrt{n(n+1)}\frac{\Delta\tau}{\alpha}\Bigr).
\tag{87}
\end{align*}

Let us note that a difference between two neighbouring energy
levels (81) can be estimated. After some appropriate series expansions,
which are valid at $n> 1$, one obtains:
\begin{align*}
&E_{n+1}-E_n \approx\frac{1}{\alpha}
\Bigl[1+\frac18\,\frac{1}{(n+1)^2}+\frac{7}{128}\,\frac{1}{(n+1)^4}\dots \Bigr]
\\
&\qu\approx \frac{1}{\alpha}\Bigl[1+\frac{1}{8 n^2}-\frac{1}{4 n^3}+
\frac{55}{128 n^4} \dots \Bigr].
\tag{88}
\end{align*}
Right Hand Side of (88) demonstrates that the levels (81) can
approximately be treated as equi-distant provided the inverse
powers of $n$ are neglected in (88), say, for the values $n>
n_0=10$. In its turn, the following series expansion is valid:
$$
\frac{n+1/2}{\sqrt{n(n+1)}}
= 1+\frac1{8 n^2}-\frac1{8 n^3}+\frac{15}{128 n^4}- \dots
\eqno{(89)}
$$
It is remarkable that the terms $\sim n^{-1}$ are absent both in
(88) and (89). Let us remind that leading asymptotical estimations,
which are obtainable with so-called {\it logarithmic accuracy}, are
important for physical applications. As it will be clear below,
the problem at hands just admits an estimation with leading
logarithmic accuracy. In this case, the inverse powers of $n$ can
be omitted with the same accuracy in (88) and (89) at $n> n_0=10$.
For such approximation, the energy levels (81) turn out to be
equi-distant, while the corresponding ratio
$\frac{n+1/2}{\sqrt{n(n+1)}}$ in (87) becomes equal to unity.
Convergency of the series (87) is not affected in this situation,
while the term omitted can be estimated.

Let us consider the exponent in (87):
$$
\sqrt{n(n+1)}\frac{\Delta\tau}{\alpha}=
\frac{\Delta\tau}{\alpha}\Bigl( n+\frac12\Bigl)-
\frac{\Delta\tau}{\alpha}\Bigl(\frac{1}{8 n}-\frac{1}{16 n^2}+
\frac{5}{128 n^3}- \dots\Bigr).
\eqno{(90)}
$$
The second term in (90) can also be neglected in the exponent at
$n> n_0$, provided $\Delta\tau/\alpha\ll 1$. This means that the
series entering into (87) can approximately be written as follows:
\begin{align*}
\sum\limits_{n=1}^{n_0}\,\frac{n+1/2}{\sqrt{n(n+1)}}\,
&P_n\Bigl(\frac{x}{R_c}\Bigr)
P_n\Bigl(\frac{x^{\prime}}{R_c}\Bigr)\,
\exp\Bigl(-\sqrt{n(n+1)}\frac{\Delta\tau}{\alpha}\Bigr)
\\
& + e^{-\Delta\tau/(2 \alpha)}\sum\limits_{n=n_0+1}^{\infty}\,
P_n\Bigl(\frac{x}{R_c}\Bigr)
P_n\Bigl(\frac{x^{\prime}}{R_c}\Bigr)\,
\Bigl(e^{-\Delta\tau/\alpha}\Bigr)^n, \tag{91}
\end{align*}
where $n_0$ is the number, which is fixed (its specific value is
forbidden to go to infinity). The correction, say $\widetilde c$,
omitted in the representation (91) can be estimated:
$$
\aligned
|\widetilde c| &\le\Big(\Bigl(1 + \frac{1}{n_0^2}\Bigr)
\Bigl(1 + \frac{\Delta\tau}{\alpha}\frac{\const}{n_0}\Bigr)-1 \Big)
\\
&\qu\times e^{-\Delta\tau/(2 \alpha)}\,\sum\limits_{n=n_0+1}^{\infty}\,
\Big|P_n\Bigl(\frac{x}{R_c}\Bigr) P_n\Bigl(\frac{x^{\prime}}{R_c}\Bigr)
\Big|\,\Bigl(e^{-\Delta\tau/\alpha}\Bigr)^n.
\endaligned
$$
It can also be demonstrated that absolute value of the first term
in (91) does not exceed $2 n_0$. Using (91), let us put (87) in
the following form:
{\small
\begin{align*}
&G(x, \tau ; x^{\prime},\tau^\prime) \approx \frac{-g \beta}{4
R_c}
\Big[\Bigl(\frac12-\frac{\Delta\tau}{\beta}\Bigr)^2-\frac1{12}\Big]
\\
&\qq-\frac{g}{2\hbar v} \sum\limits_{n=1}^{n_0}\,\Big[
\frac{n+1/2}{\sqrt{n(n+1)}}
\exp\Bigl(-\sqrt{n(n+1)}\frac{\Delta\tau}{\alpha}\Bigr)
-\exp\Bigl(-\Big(n+\frac12\Big)\frac{\Delta\tau}{\alpha}\Bigr)\Big]
\\
&\hphantom{\qq-\frac{g}{2\hbar v} \sum\limits_{n=1}^{n_0}\,}\times P_n\Bigl(\frac{x}{R_c}\Bigr) P_n\Bigl(\frac{x^{\prime}}{R_c}\Bigr)
\\
&\qq-\frac{g}{2\hbar v}\,e^{-\Delta\tau/(2 \alpha)}\Big[
\sum\limits_{n=0}^{\infty}\,t^n\,
P_n\Bigl(\frac{x}{R_c}\Bigr) P_n\Bigl(\frac{x^{\prime}}{R_c}\Bigr)-1\Big],
\tag*{\normalsize(92)}
\end{align*}}
\unskip where $t\equiv\exp(-\Delta\tau/\alpha)$. As it can be seen
from (80), the equation obtained (92) is valid for the case when
$\tau$ and $\tau^\prime$ are close either to zero or to $\beta$, as well
as for the case when only $\tau$ or $\tau^\prime$ is close to
$\beta$. Besides, we assume that $\tau\ne\tau^\prime$ in order to keep
convergency of (92).

Using the known series \cite{35}
$$
\sum\limits_{n=0}^\infty\,t^n P_n(\cos\vartheta_1) P_n(\cos\vartheta_2)=
\frac4\pi\,\frac{1}{u_+ u_-}\,{\Bbb K}(\kappa),
\quad
0 <t< 1,
\eqno{(93)}
$$
$$
\aligned
u_+\equiv {\sqrt{ 1-2t\cos(\vartheta_1+\vartheta_2)+t^2}},
\quad
u_-\equiv {\sqrt{1-2t\cos(\vartheta_1-\vartheta_2)+t^2}},&
\\
\kappa = \frac{u_+-u_-}{u_++u_-},&
\endaligned
\eqno{(94)}
$$
where ${\Bbb K}$ is a complete elliptic integral of the first
kind, one can estimate the approximate representation for the
Green function (92). Indeed, let us put $\cos\vartheta_1\equiv
x/R_c\ll 1$ and $\cos\vartheta_2 \equiv x^\prime/R_c\ll 1$. Then
the following estimations for $u_+$ and $u_-$ can be obtained:
\begin{align*}
&u_+\approx 1+t-\frac{t}{1+t}\,\frac{(x+x^\prime)^2}{2 R^2_c}
\approx\,2-\frac{\Delta\tau}{\alpha}-\frac{(x+x^\prime)^2}{4 R^2_c},
\tag{95.1}
\\
&u_-\approx\Bigl((1-t)^{2}+t \frac{(x-x^\prime)^2}{R^2_c}\Bigr)^{1/2}
\approx\frac{||x-x^\prime|+i \hbar v\Delta\tau|}{R_c}\equiv u_*,
\tag{95.2}
\end{align*}
where it is assumed that $\Delta\tau/\alpha\ll 1$ and
$$
\frac{t}{1+t}\approx\frac12\Bigl(1-\frac{\Delta\tau}{2\alpha}\Bigr).
$$

Provided the terms of second order smallness are neglected, an
estimation $\kappa\simeq 1 - u_*$ arises for $\kappa$ (94), where
$u_*$ implies the corresponding approximate value of $u_-$ given by
Right Hand Side of (95.2). When $\kappa\sim 1$, several leading
terms of the asymptotical expansion of the function ${\Bbb
K}(\kappa)$ \cite{35} can be written down:
$$
{\Bbb K}(\kappa)\approx {\Bbb K}(1-u_*)\approx
\frac{u_*}{4}\Big(\Bigl(\frac{2}{u_*}+1\Bigr)
\ln\frac{\,8\,}{u_*}-1\Big),
\quad
u_*\ll 1.
\eqno{(96)}
$$
The estimation presented (96) does not contain the terms $\sim
(u_*)^2$, and the other terms which contain higher powers of
$u_*$ are absent. This is due to the fact that under the
imposed condition of quasi-homogeneity, the value $u_*$ (95.2) is
treated as a quantity of first order smallness. Therefore, the
corresponding value of the argument $\kappa$, i. e., $\kappa\simeq
1-u_*$, is written neglecting the contributions of second order
smallness.

Let us point out that the first and the second terms in $G(x,
\tau; x^{\prime}, \tau^\prime)$ (92) are not small, because the number
$n_0$ can occur to be large. Besides, the inequality $\beta/\alpha
\gg 1$ implies that $g \beta/R_c \gg g/(\hbar v)$. However,
provided a smallness of the quantities $x/R_c$, $x^\prime/R_c$ and
$\Delta\tau/\alpha$ is taken into account, it is seen that the first two
terms in (92) are less important in comparison to the third one,
which can be logarithmically large at sufficiently small $u_*$. It
is why, we neglect the first two terms and write down the
leading contribution to the Green function (92) with the
logarithmic accuracy:
\begin{align*}
&G(x, \tau ; x^{\prime }, \tau^\prime)
\\
&\qu\simeq
\Bigl(\frac{-g }{4\pi\hbar v}\Bigr)\,
\Big(1+ \frac{2 R_c}{||x-x^\prime|+i \hbar v (\tau-\tau^\prime)|}\Big)
\ln \frac{8 R_c}{||x-x^\prime|+i \hbar v (\tau-\tau^\prime)|}
\\
&\qu\approx\Bigl(\frac{-g}{2\pi\hbar
v}\Bigr)\,\frac{1}{u_*}\,\ln\frac{8}{u_*}.
\tag{97}
\end{align*}
Here it is assumed that $u_*$ is given by (95.2), and the
following conditions of validity of the logarithmic estimation are
respected:
$$
n_0\lesssim\frac{1}{u_*},
\quad
1\ll\frac{1}{u_*}\ll\frac{1}{u_*}\ln \frac{1}{u_*}.
\eqno{(98)}
$$

Substitution of (97) into (52) gives us the following estimation
for the two-point correlator $\Gamma (x_1, \tau_1; x_2, \tau_2)$:
$$
\Gamma (x_1, \tau_1; x_2, \tau_2) \simeq
\frac{\sqrt{\widetilde\rho(x_1) \widetilde\rho(x_2)}}{
|| x_1-x_2|+i\hbar v(\tau_1 -\tau_2)|^{1/\bar\theta}}.
\eqno{(99)}
$$
In (99), the notation $\bar\theta$ for the critical exponent is
introduced:
$$
\bar\theta \equiv \frac{2\pi\hbar v}{g}\,u_*.
\eqno{(100)}
$$
The critical exponent $\bar\theta$ depends on $u_*$ (95.2), and
therefore it is a function of differences of the coordinates. Apart
from the inequalities (98), the following estimations can be obtained
to characterize the relations (99), (100):
$$
\aligned
&\frac{1}{\hbar v}\ll \frac{R_c}{\hbar v ||x-x^\prime|+
i \hbar v (\tau-\tau^\prime)|}\ll \frac{\beta}
{||x-x^\prime|+i \hbar v (\tau-\tau^\prime)|},
\\
&\frac{1}{\hbar v}\ll \frac{\beta}{R_c}
\ll \frac{\beta}{||x-x^\prime|+i \hbar v (\tau-\tau^\prime)|}.
\endaligned
$$

The estimation obtained (99) (together with the critical exponent
$\bar\theta$ (100)) constitutes the main result of the present
subsection devoted to the case given by $k_BT\ll \hbar v/R_c$. From a
comparison with the spatially homogeneous Bose gas, one can see that
the derivation of the estimate (99) is just analogous to a transition from
the relations (56), (58) to the final asymptotics (59). Then, validity of
the corresponding limit $R_c/(\hbar\beta v)\to 0$ means that the
result (99) is also concerned with the fact that the condensate
boundary $R_c$ increases slower than the inverse temperature.

Let us note that due to (98), a specific value of $n_0$ can be
related to the size of the trap $R_c$: at a fixed range of
deviations between the spatial coordinates $x$ and $x^\prime$,
increasing of $R_c$ leads to an increasing of an upper bound for
admissible values of $n_0$. However, due to (98), the estimation
obtained for $G(x, \tau ; x^{\prime }, \tau^\prime)$ (97) does not
depend explicitly on a specific choice of the number $n_0$. On the
other hand, at sufficiently large values of $n$, the following
asymptotics for the Legendre polynomials $P_n$ is valid \cite{35}:
$$
P_n (\cos\vartheta) =
\sqrt{\frac{2}{\pi n\sin\vartheta}}\,\cos\Big[
\Bigl(n+\frac12\Bigr)\vartheta-\frac\pi 4\Big]+O (n^{-3/2}),
\quad
0<\vartheta <\pi.
\eqno{(101)}
$$
Let us assume that the number $n_0$ is large enough to substitute
(101) into the series
$$
\sum\limits_{n=1}^{\infty} \Bigl(e^{-\Delta\tau/\alpha}\Bigr)^n
P_n\Bigl(\frac{x}{R_c}\Bigr)
P_n\Bigl(\frac{x^{\prime}}{R_c}\Bigr),
$$
which is a part of the representation (92) (see also (91)), in order
to obtain its estimation with the logarithmic accuracy.
Eventually, the following result takes place for $G(x, \tau ;
x^{\prime }, \tau^\prime)$ (92):
$$
G(x, \tau ; x^{\prime }, \tau^\prime) \simeq
\Bigl(\frac{-g}{2\pi\hbar v}\Bigr)\,
\Bigl(1+\frac{S^2}{2 R_c^2}\Bigr)
\,\ln\frac{R_c}{||x-x^\prime|+i \hbar v(\tau-\tau^\prime)|},
\eqno{(102)}
$$
where $S=\frac{x_1+x_2}2$. In the limit $1/R_c\to 0$, the total
coefficient in front of the logarithm in (102) takes the value
$-1/\theta$, where the critical exponent $\theta$ is defined like in
(59), (60).

The usage of the asymptotics (101) implies that the eigen-functions are
approximated by the base consisting of the plane waves, which correspond
to an almost homogeneous Bose gas. Therefore, a replacement of the
asymptotics (97) by the asymptotics (102) can be explained as a
transition, at increasing of $R_c$, to smaller scales characterized by
small ratios $x/R_c$, $x^\prime/R_c$ and $|x-x^\prime|/R_c$ (the
quasi-homogeneity condition). Then, the result (99) together with the critical
exponent (100) demonstrate an effect of finitness of the size of the
domain occupied by the spatially non-homogeneous Bose gas. This follows
just from the employment of the Legendre polynomials as the base of
one-particle states.

\head
\S5.~Conclusion
\endhead

The model considered in this paper describes a spatially
non-homogeneous one-dimensional Bose gas with a weak repulsive
coupling placed into an external harmonic potential. The paper
deals with an application of the functional integration approach
to the calculation of the two-point thermal correlation function
of the non-homogeneous Bose gas. The temperatures that are low
enough for the quasi-condensate to be created in the Bose system
in question (see~\S2) are studied. The functional integral
representation for the two-point correlation function in question
is estimated by means of the stationary phase approximation in the
way proposed in \cite{25}. The main results are obtained for the
case when the size of the domain occupied by the quasi-condensate
increases, while the temperature of the system goes to zero. It is
demonstrated that the behaviour of the correlation function near
zero temperature has a power-like dependence, and it is governed
by the critical exponent. In contrast with the case of spatial
homogeneity of the Bose gas, the presence of the external
potential is manifested in the non-homogeneity of the critical
exponent. The latter turns out to become a function of the same
spatial and thermal arguments as the correlator itself. The
dependence of the critical exponent on these spatial arguments is
in correspondence with the limiting behaviour of the ratio of the
size of the trap to the inverse temperature provided both of them
are increasing.

\subhead{Acknowledgements}
\endsubhead
The present paper has been supported in part by RFBR (the project
04--01--00825) and by the programme of Russian Academy of Sciences
``Mathematical Methods in Non-Linear Dynamics''.

\end{document}